\documentclass[11pt]{article}
\usepackage{hyperref}
\usepackage{authblk}
\pdfoutput=1

\usepackage{hyperref}
\pdfoutput=1

\begin{document}
\title{Inertial particle entrainment in a shearless mixing layer}
\author[1]{Peter J. Ireland}
\author[2]{Martin S. Henke}
\author[1]{Lance R. Collins}

\affil[1]{Cornell University, Ithaca, NY, USA}
\affil[2]{Technische Universit\"{a}t M\"{u}nchen, Germany}

\maketitle
\begin{abstract}
This article accompanies the submission of a fluid dynamics video (entry V032) of inertial particle entrainment in a shearless mixing layer for the Gallery of Fluid Motion of the 64th Annual Meeting of the American Physical Society, Division of Fluid Dynamics.
\end{abstract}

\section{Numerical simulation in video}
The video shows isocontours of the turbulent kinetic energy (left) and the concentration of inertial particles (right) in a shearless turbulent mixing layer with a step change in turbulent kinetic energy. The data is from a direct numerical simulation in a periodic cube using $512^3$ grid points.
One quarter of the cube is initialized with high turbulence levels, and the remainder of the cube is initialized with low turbulence levels.  The ratio of the turbulent kinetic energy between the two regions is approximately 30. 
The initial Taylor microscale Reynolds number in the high turbulence region is 111.
Approximately 8 million inertial particles are placed uniformly in the central, high-turbulence region of the cube, 
with the same Stokes number distribution as in the experiments of [1].
The flow evolves for a total of 7000 time steps, or approximately 2.5 large-eddy turnover times, with frames recorded every 10 time steps.

The particle concentration isocontours are calculated by binning the particles into $128^3$ boxes and computing the density of particles in each box. The concentration levels are scaled by the particle density in the seeded region, and the colorbar represents concentrations ranging linearly from 0 (blue) to 1 (red).
The turbulent kinetic energy is scaled by the average turbulent kinetic energy in the high energy region at initialization, and the colorbar ranges linearly from 0 (blue) to 1 (red).

As the flow evolves, large, energetic eddies mix particles from the seeded region to the non-seeded regions. 
As a result, the particles in the non-seeded regions are strongly clustered near the large eddies.
The low turbulence in the non-seeded regions is not very effective at mixing these particle clusters, 
and we see that this clustering persists throughout much of the video.
Near the initial interface between high and low turbulence, pockets of high turbulence penetrate what was originally a low turbulence region. This mixing region is therefore marked by high levels of intermittency.

\section{Physical relevance}
This mixing process is analagous to that which occurs in cumulus clouds, where highly turbulent clouds laden with water droplets (i.e., inertial particles) mix with the droplet-free ambient environment which is marked by much lower turbulence levels. 
A detailed examination of the physical processes involved is given in [2].

\section{Acknowledgements}
This work was supported by the National Science Foundation through CBET grants 0756510 and 0967349 and through a graduate research fellowship to Peter Ireland.  The simulations and visualizations were performed on TeraGrid resources provided by the Texas Advanced Computing Center (TACC Ranger and TACC Spur) under grants TG-CTS100052 and TG-CTS100062.

\section{References}
[1] {\sc Gerashchenko, S., Good, G. \& Warhaft, Z.} 2011 Entrainment and mixing of water droplets across a shearless turbulent interface with and without gravitational effects. \emph{J. Fluid Mech.} {\bf 668}, 293--303.\\~
[2] {\sc Ireland, P. J. and Collins, L. R.} 2011 Direct numerical simulation of inertial particle entrainment in a shearless mixing layer. \emph{J. Fluid Mech.}, in preparation.

\end{document}